\documentclass{article}
\usepackage{ijcai20}

\usepackage{times}
\usepackage{soul}
\usepackage{url}
\usepackage[hidelinks]{hyperref}
\usepackage[utf8]{inputenc}
\usepackage[small]{caption}
\usepackage{graphicx}
\usepackage{amsmath}
\usepackage{amsthm}
\usepackage{booktabs}
\usepackage{algorithm}
\usepackage{algorithmic}
\usepackage{float} 
\usepackage{subfigure}
\usepackage{booktabs}  
\usepackage{threeparttable}  
\usepackage{multirow}
\usepackage{enumitem}
\usepackage{bm}

\title{AutoAlpha: an Efficient
Hierarchical Evolutionary Algorithm for Mining
Alpha Factors in Quantitative Investment}

\author{
    Tianping Zhang\and
    Yuanqi Li\and
    Yifei Jin\and
    Jian Li
    \affiliations
    Institute for Interdisciplinary Information Sciences (IIIS), Tsinghua University, China
    \emails
    ztp18@mails.tsinghua.edu.cn, \{timezerolyq, yfjin1990\}@gmail.com, %lijian83@mails.tsinghua.edu.cn
}

\begin{document}

\maketitle

\begin{abstract}
The multi-factor model is a widely used model in  quantitative investment. The success of a multi-factor model is largely determined by the effectiveness of the alpha factors used in the model. This paper proposes a new evolutionary algorithm called {\em AutoAlpha} to automatically generate effective formulaic alphas from massive stock datasets. Specifically, first we discover an inherent pattern of the formulaic alphas and propose a hierarchical structure to quickly locate the promising part of space for search. Then we propose a new Quality Diversity search based on the Principal Component Analysis ({\em PCA-QD}) to guide the search away from the well-explored space for more desirable results. Next, we utilize the warm start method and the replacement method to prevent the premature convergence problem. Based on the formulaic alphas we discover, we propose an ensemble learning-to-rank model for generating the portfolio. The backtests in the Chinese stock market and the comparisons with several baselines further demonstrate the effectiveness of {\em AutoAlpha} in mining formulaic alphas for quantitative trading. 
\end{abstract}

\section{Introduction}

Predicting the future returns of stocks is one of the most challenging tasks in quantitative trading. Stock prices are affected by many factors such as company performances, investors' sentiment, and new government policies, etc. To explain the fluctuation of stock markets, economists have established several theoretical models. Among the most prominent ones, the Capital Asset Pricing Model (CAPM) \cite{sharpe1964capital} dictates that the expected return of a financial asset is essentially determined by one factor, that is the market excess return, while the Arbitrage Pricing Theory (APT) \cite{ross2013arbitrage} models the return by a linear combination of different risk factors. Since then, several multi-factor models have been proposed and numerous such factors (also called abnormal returns) have been found in the economics and finance literature. For example, the celebrated Fama-French Three Factor Model \cite{fama1993common} discovered three important factors that can explain almost 90\% of the stock returns\footnote{https://en.wikipedia.org/wiki/Fama\%E2\%80\%93French\_three-factor\_model}. In quantitative trading practice, designing novel factors that can explain and predict future asset returns are of vital importance to the profitability of a strategy. Such factors are usually called {\em alpha factors}, or {\em alphas} in short. 

In 2016, the quantitative investment management firm, WorldQuant, made public 101 formulaic {\em alpha factors} in \cite{kakushadze2016101}. Since then, many quantitative trading methods have used these formulaic alphas for stock trend prediction \cite{chen2019investment}. A formulaic alpha, as the name suggests, is a kind of alpha that can be presented as a formula or a mathematical expression.

\begin{equation*}
    \text{Alpha\#101} = (\text{close}-\text{open})/(\text{high}-\text{low})
\end{equation*}

For example, the above formulaic alpha is one of the alpha factors from \cite{kakushadze2016101} and is calculated using the open price, the close price, the highest price and the lowest price of stocks on each trading day. This alpha formula reflects the momentum effect that has been observed in different market (see e.g., \cite{jegadeesh1993returns}). For each day, the alpha gives different values for different stocks. The higher the value, it is more likely that the stock will have relatively larger returns in the following days.

\begin{figure}
    \centering
    \includegraphics[height=2.2cm, width=8.5cm]{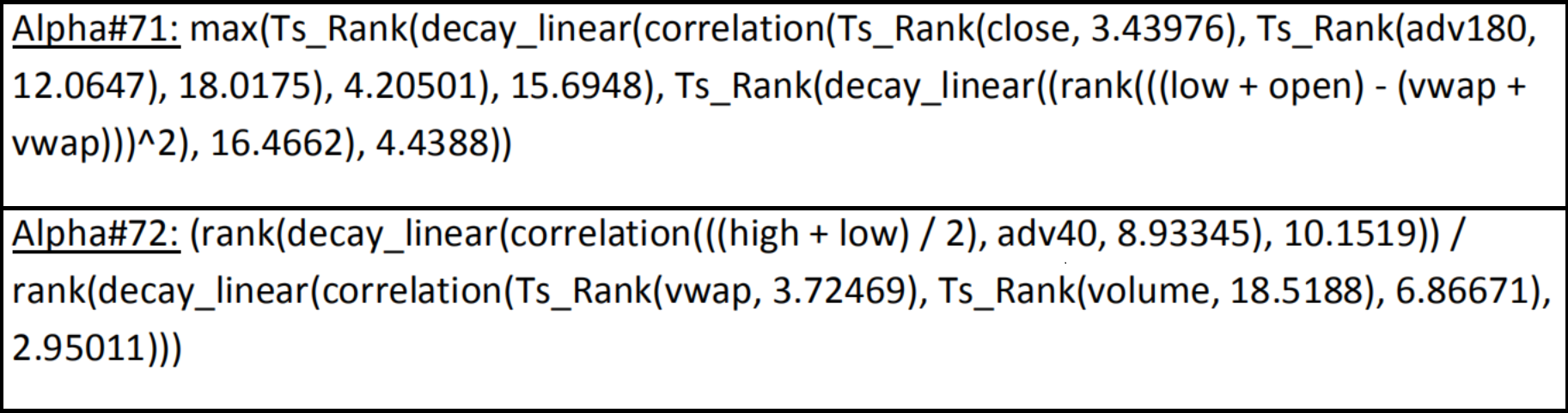}
    \caption{Formulas Alpha\#71 and Alpha\#72.}
    \label{fig:complex_alpha}
\end{figure}

There are much more complicated formulaic alphas than the one shown above. In Figure \ref{fig:complex_alpha}, we show two examples, Alpha\#71 and Alpha\#72 in \cite{kakushadze2016101}. The most common way of producing new formulaic alphas is to have economists or financial engineers to come up with new economical ideas, transform these ideas into formulas and then validate its effectiveness on the historical stock datasets. It is known that WorldQuant has been employing a large number of financial engineers and data miners (even part-time online users\footnote{https://www.weareworldquant.com/en/home}) to design new alphas. This way of finding good 
alphas requires tremendous human labor and expertise, which is not realistic for small firms or individual investors. Therefore, there is an urgent need to develop tools for mining new effective alphas from massive stock datasets automatically. 

\begin{figure}
    \centering
    \includegraphics[height=3.8cm, width=8cm]{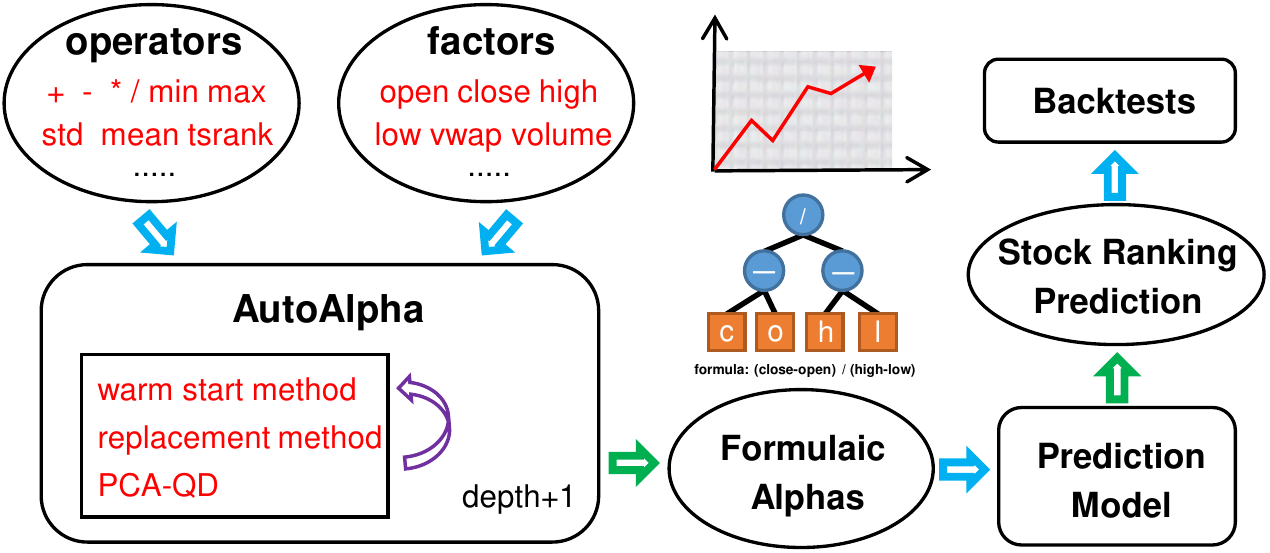}
    \caption{The framework of our approach.}
    \label{fig:procedure}
\end{figure}

Our goal is to find as many diverse formulaic alphas with desirable performance as possible within limited computational resources. Unlike many optimization and search problems which aim at finding one single desirable solution, we prefer to look for multiple diverse solutions with high performance and low correlation. 

As shown in Figure \ref{fig:crossover}, a formulaic alpha can be expressed as a tree where the leaves correspond to raw data and the inner nodes correspond to various operators. As the discrete search space is very large, it is natural to use genetic algorithms to search for effective alphas in the form of trees. However, as we argue below, this is not straightforward and there are several challenges we need to address.

\textbf{Challenge 1: Quickly locate the promising search space.} The vanilla genetic algorithm is generally inefficient in mining effective formulaic alphas, due to the fact that effective alphas are sparse in the huge search space. Therefore, how to quickly locate the promising space for search becomes a critical issue.

\textbf{Challenge 2: Guide the search away from the explored search space.} In order to find many diverse and effective formulaic alphas, we need to run the genetic algorithm several times for more results. However, the vanilla genetic algorithm usually converges to the same local minima.

\textbf{Challenge 3: Prevent the premature convergence problem.} The premature convergence problem \cite{gupta2012overview} arises in genetic algorithms when some type of effective genes dominate the whole population and destroy the diversity in the population. When premature convergence happens, the population stucks at a suboptimal state and we can no longer produce offspring with higher performance.

% Previous studies have limited efforts on addressing these problems and it remains a challenge to solve all of the above problems in a unified framework. To this end,

In this paper, we propose a new model called {\em AutoAlpha} to address the above challenges in a unified framework. 
The technical contributions of this paper can be summarized as follows:
\begin{itemize}[leftmargin= 10 pt]
    \item We discover an inherent pattern of the formulaic alphas. Based on this, we design a hierarchical structure to quickly locate the promising space for search, which address Challenge 1.
    \item For Challenge 2, we propose a new Quality Diversity method based on the Principal Component Analysis ({\em PCA-QD}) to guide the search away from the explored space for more diverse and effective formulaic alphas.
    \item We introduce the warm start method at the initialization step and the replacement method during reproduction to prevent the premature convergence problem. This addresses Challenge 3.
\end{itemize}

Based on the formulaic alphas we discover, we propose an ensemble learning-to-rank model to predict the stocks rankings and develop effective stock trading strategies. We perform backtests in the Chinese stock market for different holding periods. The backtesting results show that our method consistently outperforms several baselines and the market index.

\section{Problem Statement}
Mining formulaic alphas can be regarded as a feature extraction problem. We start from an initial set of basic factors (e.g. open, close, volume, etc.) and operators (e.g. +-*/, min, std, etc.), and then build formulaic alphas that satisfy certain performance measurement criterion, in order to reveal some inherent patterns of the stock market. The basic factors and the operators we use can be found in \cite{kakushadze2016101}. The data is public in Chinese stock markets and can be accessed through multiple resources\footnote{http://tushare.org/\#}. In this section, we formalize the problem of mining formulaic alphas.

\subsection{Stock Returns}
The return of a stock is generally determined by the close price of the stock and the holding period. For a given stock $s$, a given date $t$ and a given holding period $h$, the return of the stock can be calculated as:
{
\setlength{\abovedisplayskip}{3pt}
\setlength{\belowdisplayskip}{1pt}
\begin{equation*}
    r^{(h)}_{t,s}=\frac{close_{t+h,s}-close_{t,s}}{close_{t,s}}
\end{equation*}
}

where $close_{t,s}$ is the close price of stock $s$ at date $t$. Assuming that there are {\em n} different stocks in the stock pool, the return vector at date {\em t} of holding period {\em h} is denoted as $\bm{r}^{(h)}_{t}=(r^{(h)}_{t,1},...,r^{(h)}_{t,n})$.

\subsection{Evaluation Metrics}
For a given formulaic alpha {\em i}, its value for the stock {\em s} at date {\em t} is defined as $a^{(i)}_{t,s}$. We use the {\em IC} (information coefficient) \cite{grinold2000active} to evaluate the effectiveness of an alpha in the mining process. For a given formulaic alpha {\em i} and a given holding period {\em h}, the IC can be calculated as the mean of the {\em IC array}:
{
\setlength{\abovedisplayskip}{1pt}
\setlength{\belowdisplayskip}{1pt}
\begin{equation}
    IC_i = \frac{1}{T}\sum_{t=1}^T corr(\bm{a}^{(i)}_t, \bm{r}^{(h)}_t)
\end{equation}
}

where $\bm{a}^{(i)}_t=(a^{(i)}_{t,1},...,a^{(i)}_{t,n})$ is the value vector of formulaic alpha {\em i} at date {\em t}, $corr$ is the sample Pearson Correlation, $\{corr(\bm{a}^{(i)}_t, \bm{r}^{(h)}_t)\}_{t=1}^T$ is the IC array and {\em T} is the number of trading days in the training period.

The IC of an alpha indicates the relevance between the alpha and the stock returns, and should be as high as possible\footnote{W.l.o.g., we assume that $IC\geq 0$ since we can multiply $-1$ to a formulaic alpha which has negative IC.}.

\subsection{Similarity between Alphas}
The similarity between the alpha {\em i} and {\em j} is calculated as:
{
\setlength{\abovedisplayskip}{1pt}
\setlength{\belowdisplayskip}{4pt}
\begin{equation*}
    sim(i,j)=\frac{1}{T}\sum_{t=1}^T corr(\bm{a}^{(i)}_t,\bm{a}^{(j)}_t)
\end{equation*}
}
A group of alphas is {\em diverse} if the similarity between any two alphas in the group is lower than $0.7$.

In the process of mining formulaic alphas, our goal is to find as many {\em diverse} formulaic alphas with high IC as possible within limited computational resources.

\section{AutoAlpha}

{\em AutoAlpha} is a framework based on genetic algorithms \cite{whitley1994genetic}. Genetic algorithm is a kind of metaheuristic optimization algorithm which draws inspiration from biological process that produces new offspring and evolve a species. The vanilla genetic algorithm uses mechanisms such as reproduction, crossover, mutation and selection to give birth to new offspring. In each step of regeneration, it uses fitness function to select the best-fit individuals for reproduction. After we give birth to new offspring through crossover and mutation operations, we replace the least-fit individuals in the population with new individuals to realize the mechanism of elimination through competition. 

\begin{figure}
    \centering
    \includegraphics[height=2.2cm, width=8.2cm]{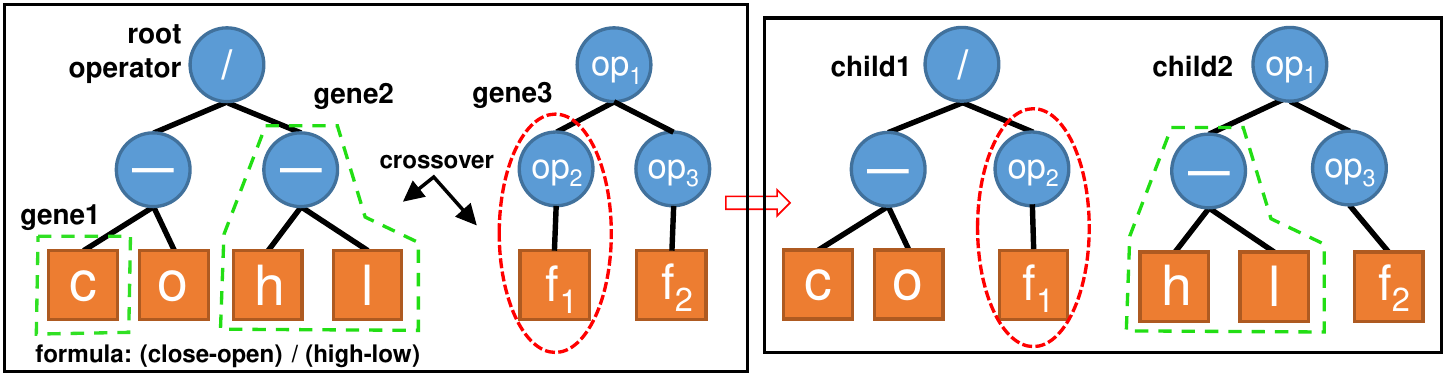}
    \caption{The demonstration of crossover. The leftmost tree shows the tree representation of the formulaic alpha '$(\text{close}-\text{open})/(\text{high}-\text{low})$'. The trees on the right are two children after crossover. op: operator. f: factor.}
    \label{fig:crossover}
\end{figure}

In order to apply the genetic algorithm for mining formulaic alphas, first we need to define the genetic representation of a formulaic alpha. As shown in the leftmost tree of Figure \ref{fig:crossover}, a formulaic alpha can be represented as a formulaic tree. It would be much easier for us to carry out crossover and mutation for trees. Figure \ref{fig:crossover} shows the crossover between two formulaic alphas of depth 2. We perform the crossover in the same depth level to prevent the depth from increasing. That is, the crossover between {\em gene1} and {\em gene3} in  Figure \ref{fig:crossover} is not allowed. The {\em gene2} and {\em gene3} are called {\em root genes} which are directly attached to their root operators while {\em gene1} is not.

\subsection{Hierarchical Structure}
The search space of trees is huge and the effective alphas are very sparse. In our experiment, we find out that the standard genetic algorithm (e.g., that implemented in python package 'gplearn') is generally inefficient in initializing the population and exploring the search space for mining formulaic alphas (see the results in Section 4.4). %It initializes the population in a random way and starts searching in a random place.  
%Motivated by some discoveries during the experiments,
For remedy, We propose a novel hierarchical search strategy for the genetic algorithm, that is significantly more
efficient in the initialization and exploration of the search space.

\subsubsection{Motivation}
In the early stage of this research, we have been using vanilla genetic algorithms for mining formulaic alphas. An interesting phenomenon occurs during the experiments that the algorithm usually converges to similar formulaic alphas with '{\em vwap/close}' as a piece of its genes. '{\em vwap}' is the Volume Weighted Average Price of a stock. The gene {\em vwap/close} itself is also an effective alpha which relates to the phenomenon of mean reversion.  While {\em vwap/close} itself is an effective formulaic alpha, the formulaic alphas of higher depth which contain {\em vwap/close} as a piece of its genes usually combine this mean reversion information with some other information and have higher effectiveness.

\begin{figure}
  \centering 
  \subfigure[depth=2]{ 
    \includegraphics[width=1.5in]{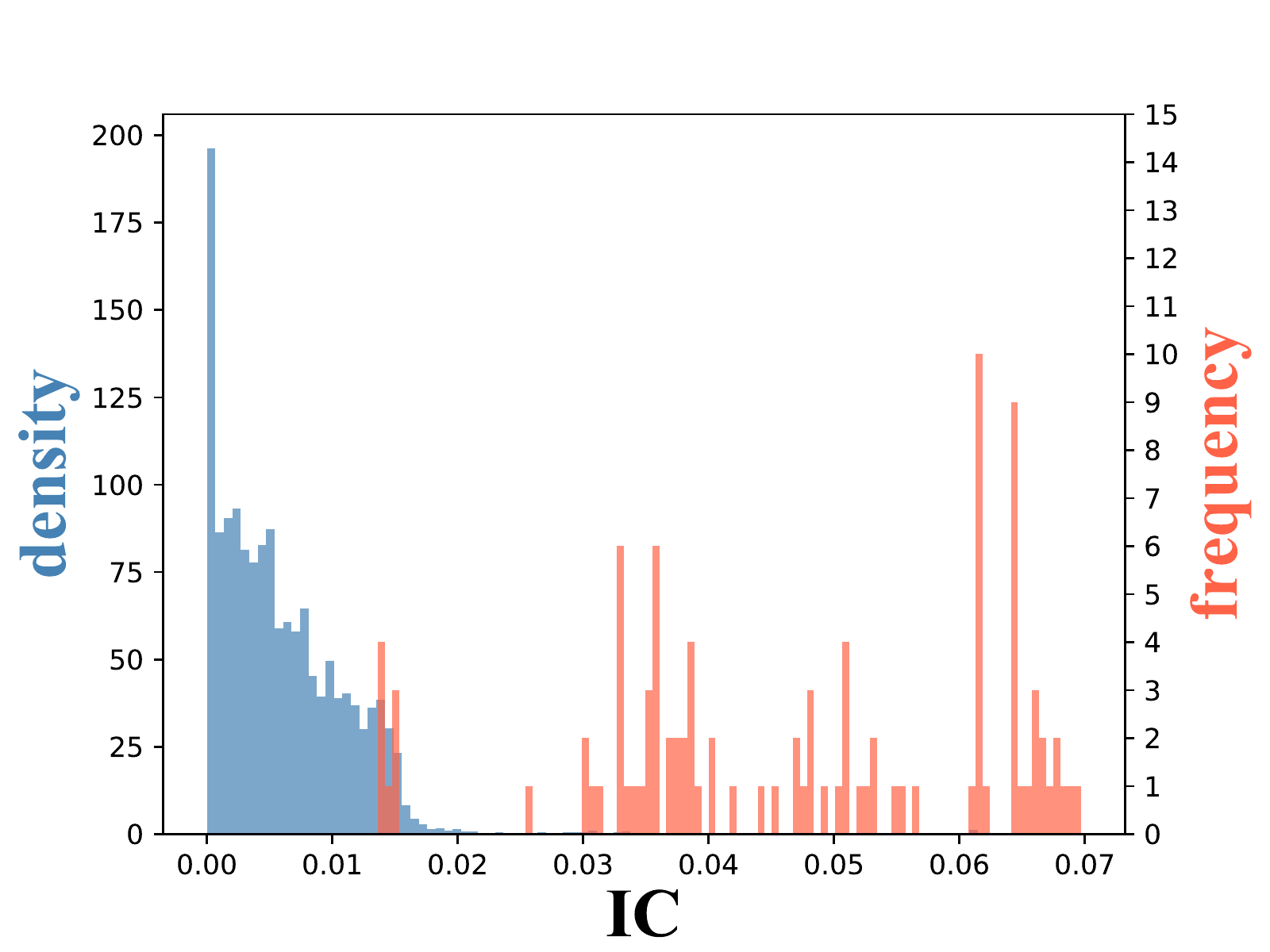} 
  } 
  \subfigure[depth=3]{ 
    \includegraphics[width=1.5in]{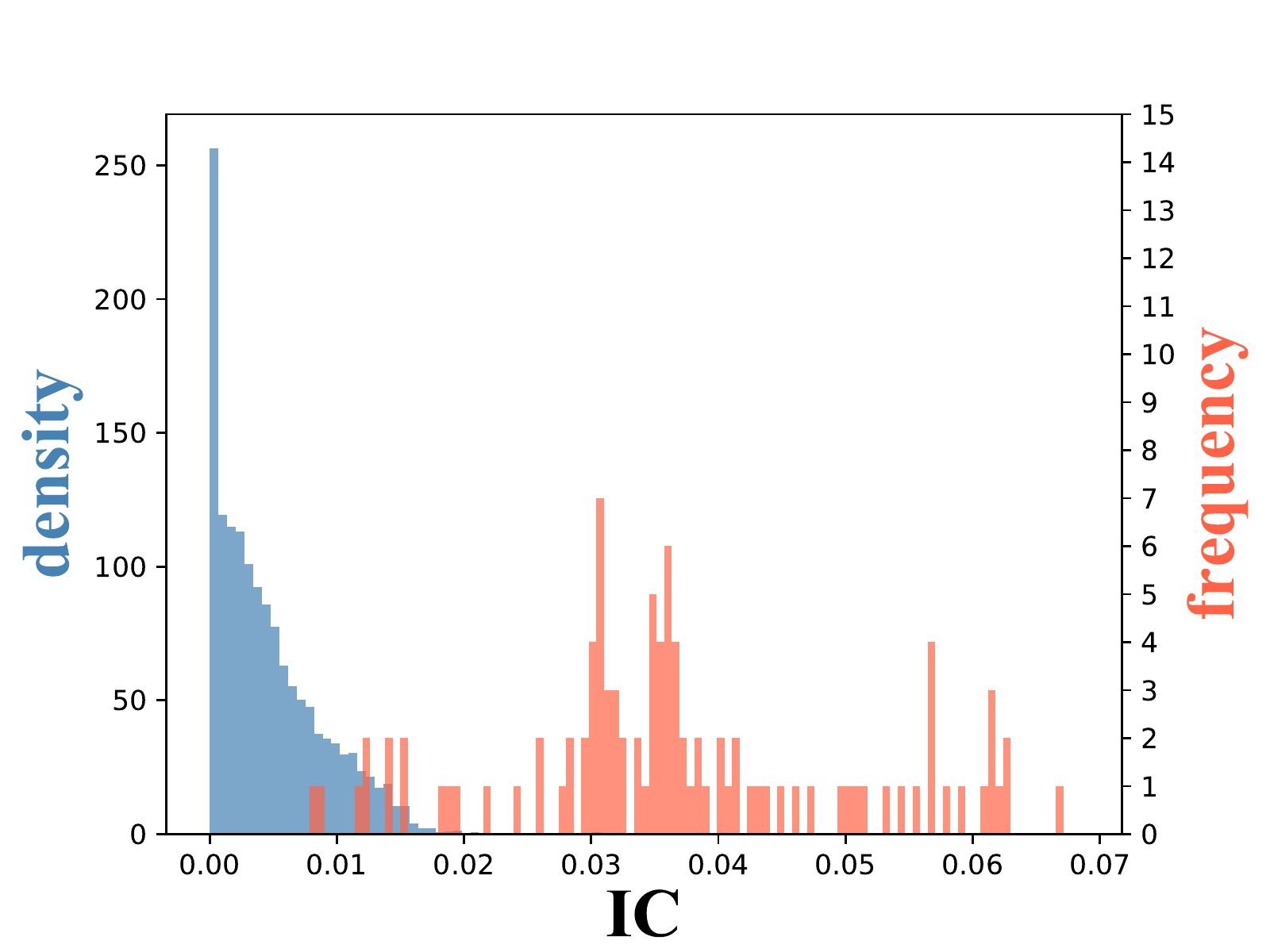} 
  } 
  \caption{The IC of the root genes of the top 100 discovered formulaic alphas and its distribution. For example, in the left figure, the blue histogram is the density plot of IC of the formulaic alphas of depth 2 (estimated by 20000 randomly generated samples). And the red histogram is the frequency plot of IC of the root genes of the top 100 discovered formulaic alphas of depth 3.} 
  \label{fig:motivation} %% label for entire Figure 
  \setlength{\belowcaptionskip}{-1cm}
\end{figure}

Based on such phenomenon, we propose a hypothesis about the inherent pattern of the formulaic alphas, that is, most of the effective alphas have at least one effective root gene. Intuitively, if we want to obtain the formulaic alphas of higher depth, we should search nearby the effective alphas of lower depth. We design an experiment to verifies the hypothesis. First, we use the vanilla genetic algorithm for evolving formulaic alphas. Then we select the top $100$ discovered formulaic alphas. For each selected alpha, we further collect its root gene with highest IC. We use the density plot to show that those root genes are effective and are hard to obtain by random generation. The results are shown in the Figure \ref{fig:motivation}.

Based on the analysis, if we maintain a population with diverse and effective root genes in it, the crossover operation attempts to searching nearby effective formulas of lower depth, and thus improve the efficiency in obtaining diverse and effective alphas. In order to establish a diverse and effective gene pool for initializing the population, we use the hierarchical structure as the framework for {\em AutoAlpha} and generate alphas from lower to higher depth iteratively.

\subsection{Guide The Search}
Now the problem has turned to how to generate formulaic alphas of each depth. Unlike many optimization and search problems which aim at finding one single desirable solution, our goal is to look for as many formulaic alphas with high IC and low correlation as possible. If we have already obtained a group of formulaic alphas, we would like to guide the search into unexplored space for more alphas. 

However, the genetic algorithms usually converge into the same local minima. In genetic algorithms community, one method to tackle such problem is the \textbf{Quality Diversity (QD)} search \cite{pugh2016quality}, which seeks to find a maximally diverse collection of individuals where each individual has desirable performance. If we can calculate the similarity (or distance) between individuals, then we can guide the search by changing the objective landscape through penalty \cite{lehman2011evolving}. For example, if the similarity between the new alpha and any of the alphas in the record exceeds a certain threshold, then the fitness of the new alpha is penalized to be $0$ (least-fit). However, calculating the similarities is slow, especially when the size of the record grows large. Assuming that the size of the record is $p$, then calculating the similarities between a new alpha and the alphas in the record is $O(npT)$ where $T$ is the number of trading days and $n$ is the number of stocks. 

In order to reduce the time complexity, we find a simpler way to approximate the similarity and design our {\em PCA-QD} search. Firstly, we use the first principal component vector to represent the information of a formulaic alpha. Specifically, the values of the formulaic alpha {\em i} can be presented as a sample matrix $\bm{A}^{(i)}=(a^{(i)}_{t,s})_{T\times n}$, where each column (stock) can be treated as a feature and each row (date) can be treated as a sample. Then we can calculate the first principal component of the sample matrix. Next, we use the Pearson Correlation between the first principal components of the two alphas, which we call {\em PCA-similarity}, to approximate the similarity between the two alphas. The calculation of the first principal component using power method is $O(nT+n^2)$. In this way, we reduce the computational complexity of calculating similarities from $O(npT)$ to $O(pT)$ (in experiments, $n$ is 300 and we have $n<T$ and $n<p$). 

We run the experiments to see how the approximation method works by random sampling. The threshold for the {\em PCA-similarity} is set to be 0.9. When the similarity is over $0.7$, the Mean Absolute Error (MAE) between the {\em PCA-similarity} and the similarity is $0.092$. And when the {\em PCA-similarity} is over $0.9$, the MAE is $0.125$. Since we are using the {\em PCA-similarity} instead of the similarity to guide the search away from the explored area, we hope that this approximation should be accurate when the similarity is high and when the {\em PCA-similarity} is high as well.

\subsection{Prevention of Premature Convergence}
The premature convergence problem has always been a critical issue in the genetic algorithms \cite{gupta2012overview}. If we always replace the least-fit individuals with new individuals of greater fitness, it is likely that some type of effective genes may dominate the whole population and destroy the genetic diversity. When premature happens, the whole population gets stuck early in a bad local minima. There are many methods addressing the premature convergence problem \cite{gupta2012overview}, and we select two from them in particular for {\em AutoAlpha}.

\textbf{Warm Start}. In the initialization step, instead of randomly generating individuals to the size of the population, we generate individuals $K$ times the size of the population and then select the individuals which rank at top $\frac{1}{K}$ according to IC into the population as initialization. In this way, we improve the average effectiveness of the initialized individuals, and thus accelerate the evolution. The warm start method in AutoAlpha helps us filter out those genes that are not useful in constructing formulaic alphas of higher depth.

\textbf{Replacement Method}. In the reproduction step, instead of comparing the new individuals with the least-fit individuals in the population, we compare the new individuals with their own parents. A pair of parents has two offspring after crossover, and the two offspring can replace their parents when the best of their fitness is greater than that of their parents. In this way, all the genes in the population only have one piece of copy after reproduction, which helps us prevent the premature convergence problem.

\subsection{Overall Algorithm of AutoAlpha}
\begin{enumerate}[leftmargin= 12 pt]
    \item We enumerate the alphas of depth 1 and select the effective ones to set up the gene pool.
    \item We use the warm start method and the gene pool to initialize the population of depth 2. Then we use crossover and the replacement method for reproduction. If the new offspring has higher IC than its parents, we will calculate its PCA-similarity with the alphas in the record and determine whether to keep it in the population
    \item We repeat step 2 for more alphas of depth 2 and update the gene pool and the record. Then we can start generating alphas of depth 3 and so forth.
\end{enumerate}

We use the formulaic alphas as input features and we use LightGBM \cite{ke2017lightgbm} and XGBoost \cite{chen2016xgboost} as learning algorithms to learn-to-rank the stocks. Then we ensemble the results and use it for stock investment in the testing period. Due to space limit, we omit the details of the training and the choice of hyper-parameters.
% More details of the training process and parameters settings are shown in the supplement.

\section{Experiments}
\subsection{Experimental Settings}
\textbf{Tasks.} We conduct the experiments independently for the holding periods of 1 day and 5 days. First we use {\em AutoAlpha} to generate a group of effective formulaic alphas for the given holding period. Then we use the ensemble model to learn to rank the stocks. Finally we use the predicted rankings to construct the stock portfolio each day and provide backtests to evaluate the effectiveness of the alphas we generate. 

\textbf{Datasets.} We use the $300$ stocks in the CSI 300 Index (hs300)\footnote{CSI 300 is a capitalization-weighted stock market index replicating the performance of top 300 stocks traded in the Shanghai and Shenzhen stock exchanges.} 
as the stock pool when mining the formulaic alphas in the training stage. We perform backtests on the stock pool of CSI 800 Index (zz800)\footnote{CSI 800 Index is comprised of all the constituents of CSI 300 Index and CSI 500
Index. It is designed to reflect the overall performance of the large, middle and
small capital stocks in Chinese stock markets.}. The overall testing period is from $20170901$ to $20190731$ and the training period is from $20100101$ to $20170831$.

\subsection{Evaluation Metrics}
We measure the performance of the trading strategies by standard metrics in the literature such as annualized return (AR), annualized volatility and Sharpe ratio (SR).

\textbf{Annualized return (AR).} Annualized return calculates the rate of return for a given holding period scaled down to a 12-month period: $\text{AR} = \mathrm{exp} \left\{365/T^{'}\times\log\left(S_{T}/S_0\right) \right\}-1$, where $T^{'}$ is the number of days and $T$ is the number of trading days. $S_t$ denotes the total wealth at the end of $t$-th trading day.

\textbf{Sharpe ratio (SR).} In finance, the Sharpe ratio is a popular measure of the risk-adjusted performance of an investment strategy:
$\text{SR}=(R_p-R_f)/\sigma_p$, where $R_p$ is the annualized return of the portfolio, $R_f$ is the risk-free rate\footnote{We set the risk-free rate in the experiments to be 0 for simplicity, since there is no consensus on the value of risk-free rate.}, and $\sigma_p$ is the annualized volatility, which is the standard deviation of the strategy's yearly logarithmic returns\footnote{https://en.wikipedia.org/wiki/Volatility\_(finance)}.

\begin{table}

\small

\centering

\caption{Performance of the top $5$ formulaic alphas for holding period $h=1,5$ in both training and testing period. The values out of/in the brackets are the results in the training/testing period.}

\label{tab:alpha_result}
\scalebox{0.8}{
\begin{tabular}{cccccc}

\toprule

\multirow{2}{*}{Top-$k$ Alpha} & \multicolumn{2}{c}{$h=1$} & \multicolumn{2}{c}{$h=5$} \\

\cmidrule(r){2-3} \cmidrule(r){4-5} 

&  hs300      &  zz800 

&  hs300      &  zz800    \\

\midrule

$Top1 $   &8.36\%(7.10\%)    & 8.41\%(7.47\%)     &8.47\%(6.15\%)    & 8.82\%(5.53\%)   \\

$Top2 $   &8.30\%(6.07\%)    & 8.17\%(3.28\%)    &7.87\%(6.52\%)    & 7.80\%(5.46\%)    \\

$Top3 $   &7.84\%(5.55\%)    &7.24\%(5.31\%)      &7.69\%(5.61\%)    &7.90\%(4.48\%)  \\

$Top4 $   &7.84\%(6.64\%)    &7.64\%(6.79\%)      &7.67\%(7.30\%)    &7.88\%(6.43\%)   \\

$Top5 $   &7.74\%(6.66\%)    & 8.02\%(5.98\%)     &7.50\%(5.53\%)    & 7.40\%(4.35\%)   \\

\bottomrule

\end{tabular}
}
\end{table}

\begin{table}

\small

\centering

\caption{Comparison with gplearn and Alpha101. $n$: number of diverse formulaic alphas with IC higher than $0.05$. $avgIC$: average IC of the top 50 discovered formulaic alphas.}

\label{tab:alpha_comparison}
\scalebox{1}{
\begin{tabular}{cccccc}

\toprule

\multirow{2}{*}{Method} & \multicolumn{2}{c}{$h=1$} & \multicolumn{2}{c}{$h=5$}  \\

\cmidrule(r){2-3} \cmidrule(r){4-5}

&  $n$      &  $avg IC$ 

&  $n$      &  $avg IC$    \\

\midrule

Alpha101     & 0   & 1.02\%     & 0   & 1.25\% \\

gplearn   & 35   & 6.10\%     & 7   & 3.35\%   \\

AutoAlpha  & 434   & 7.50\%     & 415   & 6.71\%    \\

\bottomrule

\end{tabular}
}

\end{table}

\subsection{Baselines for Comparison}
To further demonstrate the effectiveness of our method, we compare it with the following baselines:

\begin{itemize}[leftmargin= 10 pt]
    \item \textbf{Alpha101}: We compare the formulaic alphas we automatically discover with the 101 
    formulaic alphas in \cite{kakushadze2016101}. We use the 101 alphas to train the models and perform backtests under the same settings.
    \item \textbf{gplearn}: 'gplearn' is a popular python package which performs the standard genetic algorithm. It can also be used to generate formulaic alphas. 
    %We compare {\em AutoAlpha} with those typical genetic %algorithms.
    \item \textbf{SFM}: \cite{zhang2017stock} proposed SFM (the State Frequency Memory recurrent network) which is an end-to-end deep learning method and applied it to the stock prediction tasks.
    \item \textbf{Market}: The market is represented by the CSI 800 Index. We show that our method is able to outperform the market signficantly.
\end{itemize}

\subsection{Results of Generated Formulaic Alphas}
\begin{figure}[htbp]
\centering
\subfigure[Alpha1]{
\includegraphics[width=3.8cm]{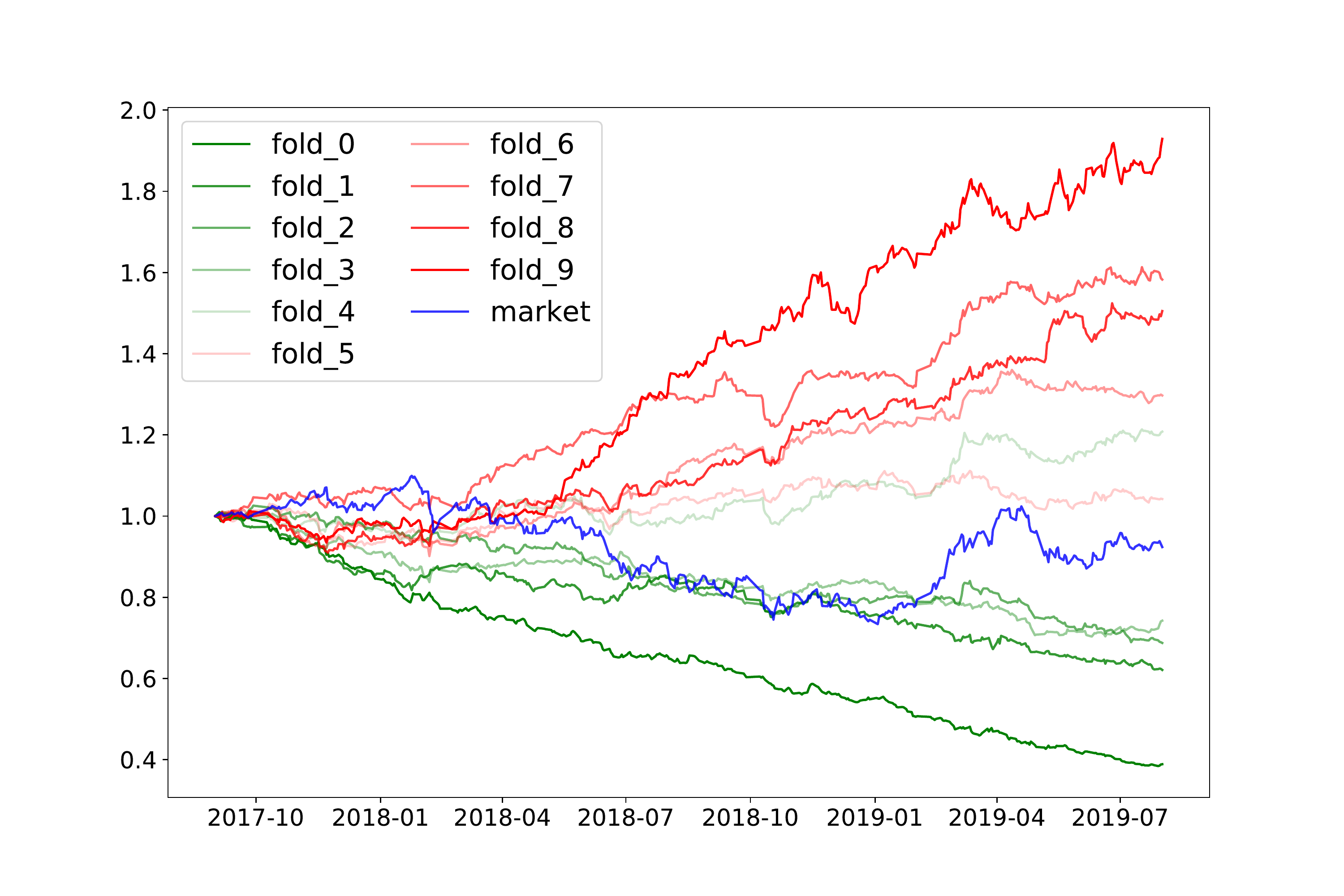}
%\caption{fig1}
}
\;
\subfigure[Alpha2]{
\includegraphics[width=3.8cm]{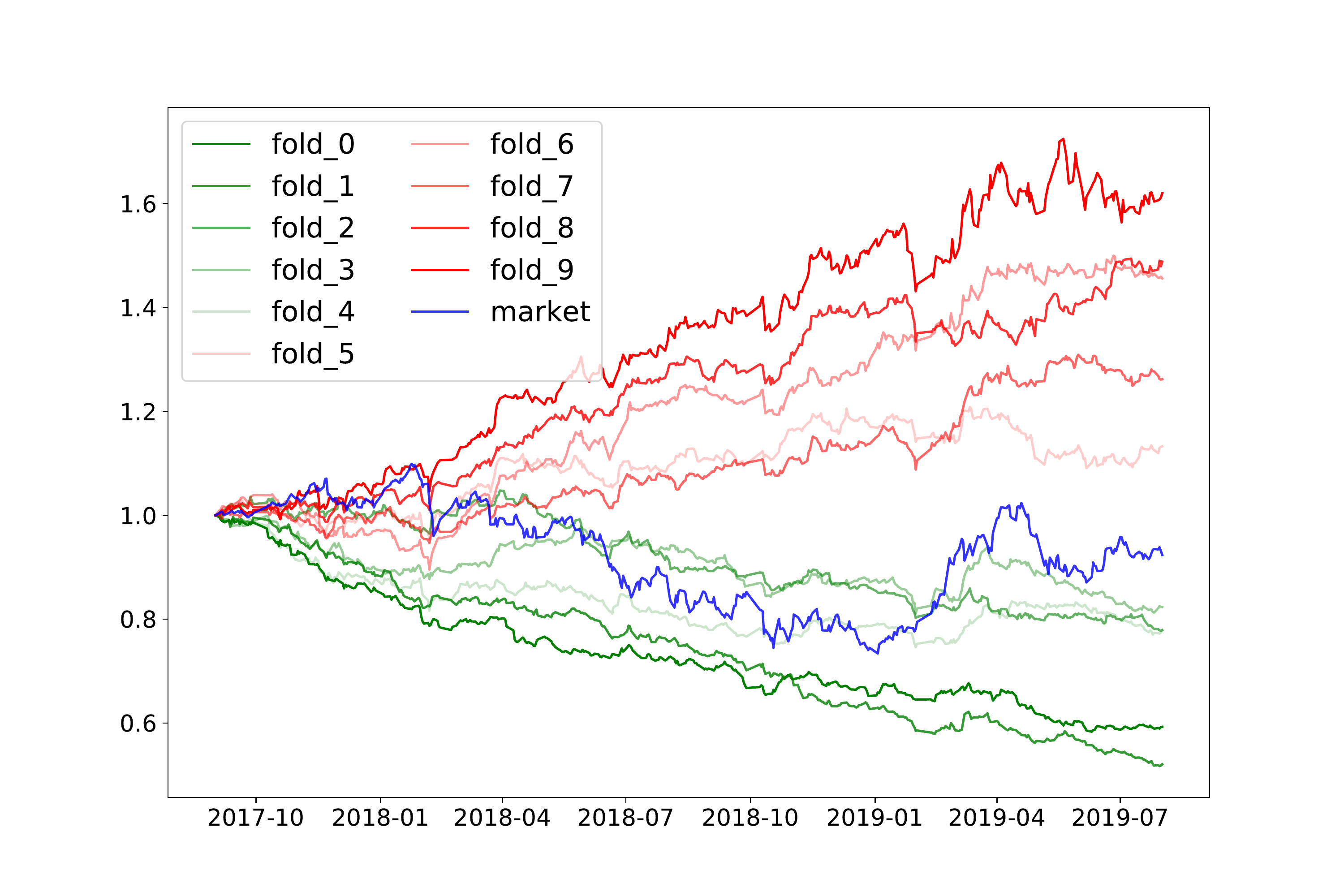}
}

\subfigure[Alpha3]{
\includegraphics[width=3.8cm]{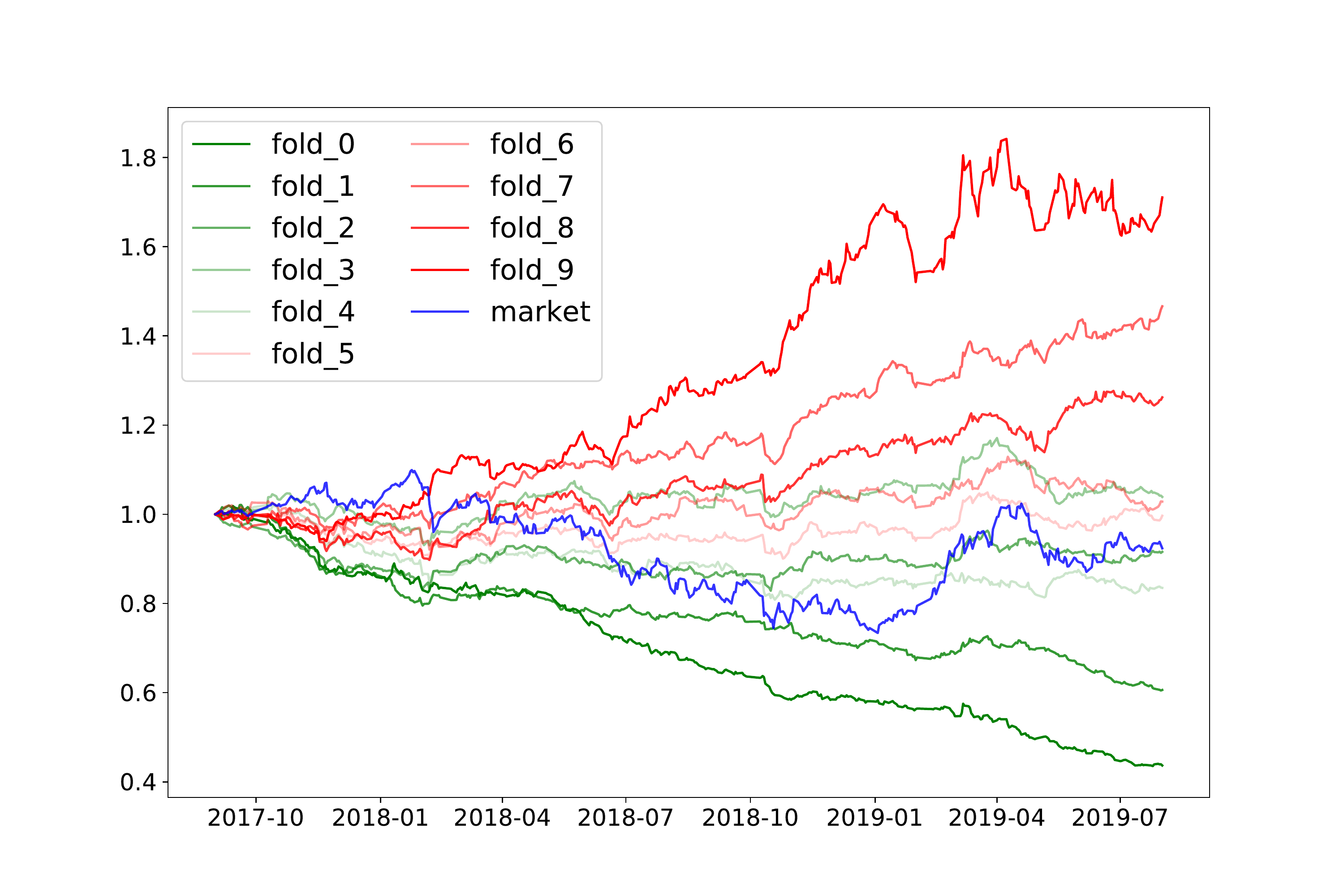}
}
\;
\subfigure[Alpha4]{
\includegraphics[width=3.8cm]{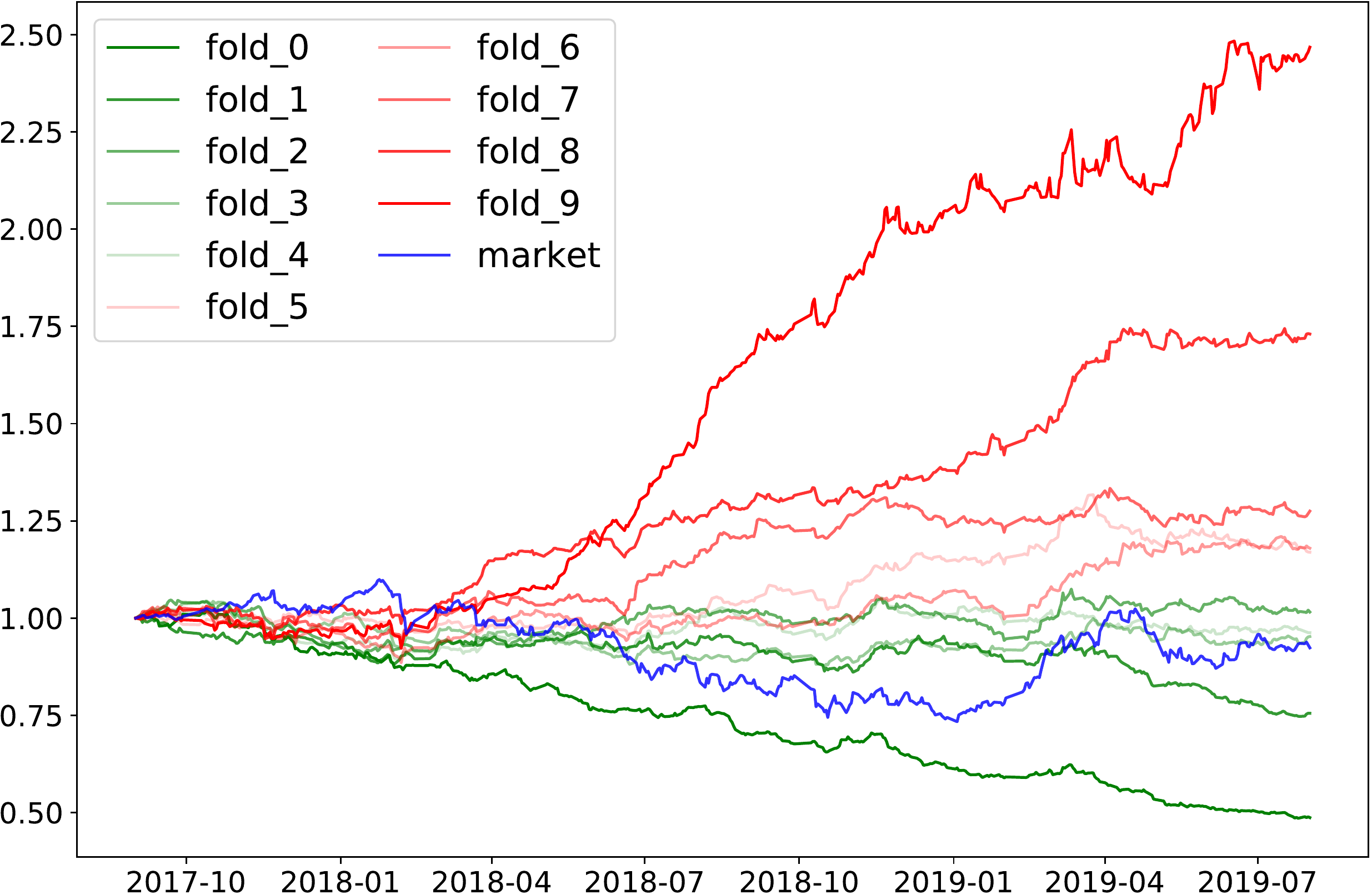}
}
\caption{The results of stratified backtests of the top 4 alphas (h=1).}
\label{fig:stratified_back-testing}
\end{figure}

Table \ref{tab:alpha_result} shows the IC of the top 5 generated formulaic alphas. We show that the alphas can not only generalize in the testing period, but also generalize in the different stock pool of CSI 800 Index. Table \ref{tab:alpha_comparison} shows the comparison between the alphas generated by {\em AutoAlpha}, the alphas generated by gplearn and the 101 alphas in \cite{kakushadze2016101}. We use two metrics to compare the results from both quantitative and qualitative aspects. One is the number of diverse formulaic alphas with IC higher than $0.05$. Another is the average IC of the top $50$ diverse formulaic alphas. We use the stratified backtests to show the alphas' capability in ranking stocks. The stock pool is divided equally into $10$ fold each day according to the ratings the alpha gives and fold 9 has the stocks with highest ratings. Then the $i$-th strategy always buy the stocks in the $i$-th fold each day. The results are shown in Figure \ref{fig:stratified_back-testing}.

\begin{table}

\small

\centering

\caption{Results of backtests and comparisons with baselines. Results relative to the market are in the brackets.}

\label{tab:back-testing}
\scalebox{0.8}{
\begin{tabular}{cccccccc}

\toprule

\multirow{2}{*}{Method} & \multicolumn{2}{c}{$h=1$} & \multicolumn{2}{c}{$h=5$}  \\

\cmidrule(r){2-3} \cmidrule(r){4-5}

&  $AR$      &  $SR$   &  $AR$      &  $SR$  \\

\midrule

market & -4.1\%  & -0.20  & -6.4\%  & -0.31 \\

SFM  & -60.0\%(-58.5\%)   & -2.05(-2.89\%)       & -23.7\%(-17.7\%)   & -1.01(-1.53)    \\

gplearn   & 61.8\%(68.7\%)   & 2.34(4.26)      & 16.9\%(22.6\%)   & 0.72(1.93)   \\

Alpha101      & 29.5\%(35.3\%)   & 1.06(2.02\%)     & 16.3\%(23.3\%)   & 0.70(2.09)  \\

our method & \textbf{90.0\%(98.2\%)} & \textbf{3.39(6.02)}  & \textbf{28.0\%(34.0\%)}  & \textbf{1.20(3.05)}\\

\bottomrule

\end{tabular}
}

\end{table}

\subsection{Backtesting Results}

\begin{figure}[H] 
  \centering 
  \subfigure[holding period $h=1$]{ 
    \includegraphics[height=3cm, width=8cm]{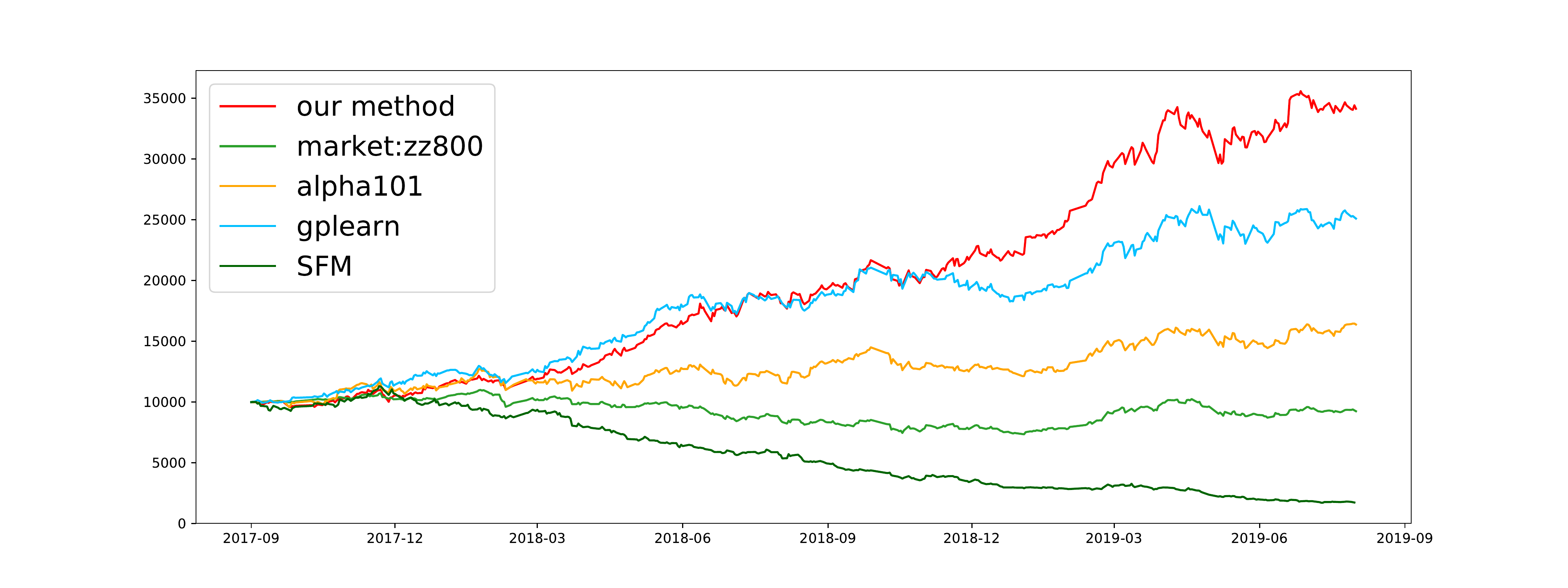} 
  } 
  \subfigure[holding period $h=1$ (relative to the market)]{ 
    \includegraphics[height=3cm, width=8cm]{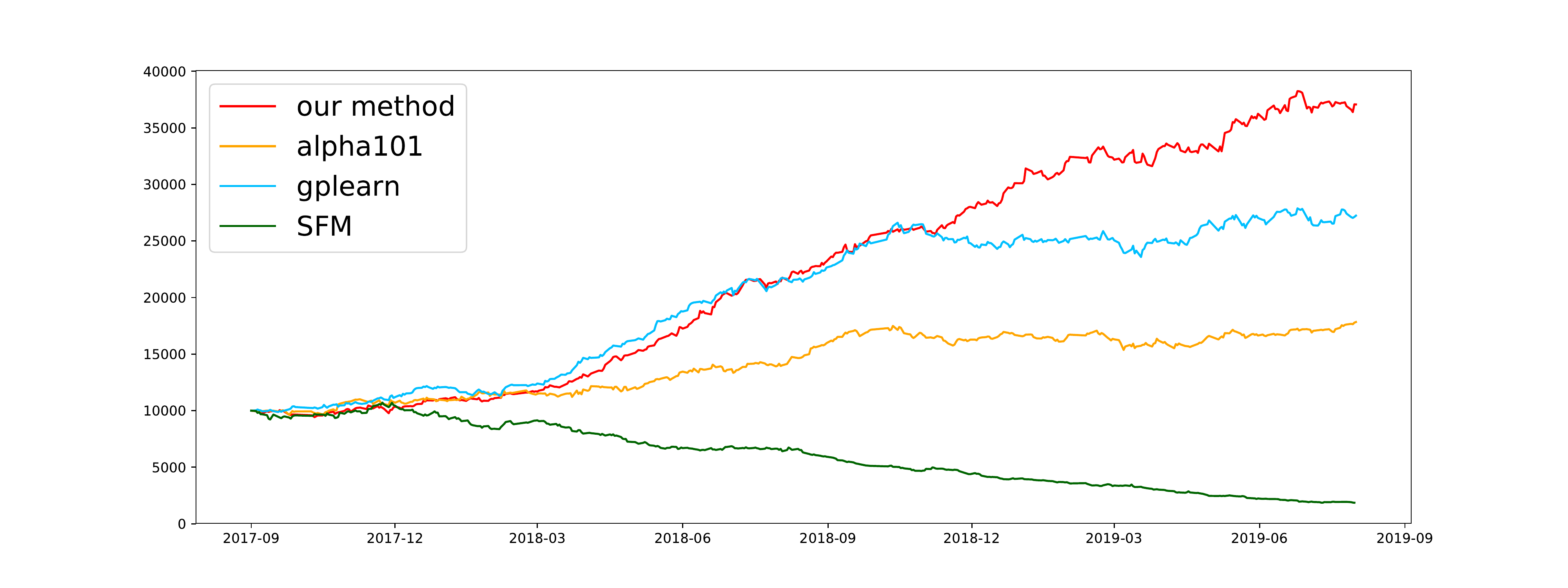} 
  } 
  \subfigure[holding period $h=5$]{ 
    \includegraphics[height=3cm, width=8cm]{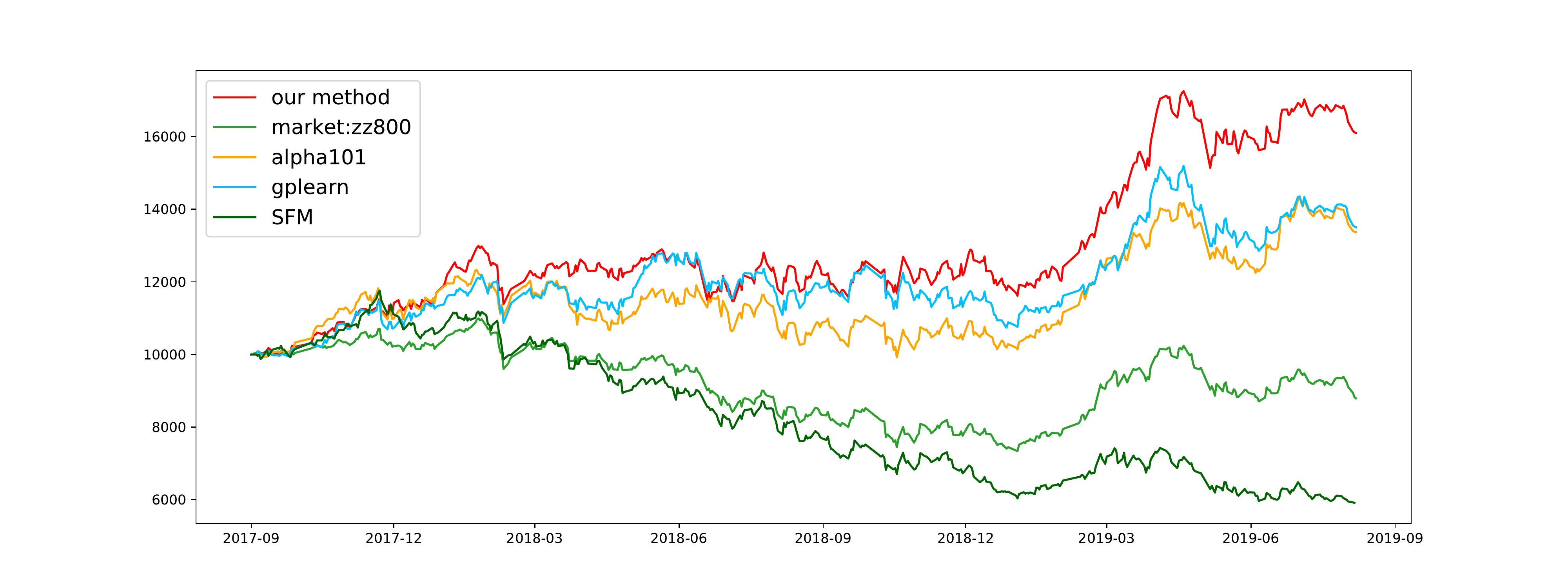} 
  } 
  \subfigure[holding period $h=5$ (relative to the market)]{ 
    \includegraphics[height=3cm, width=8cm]{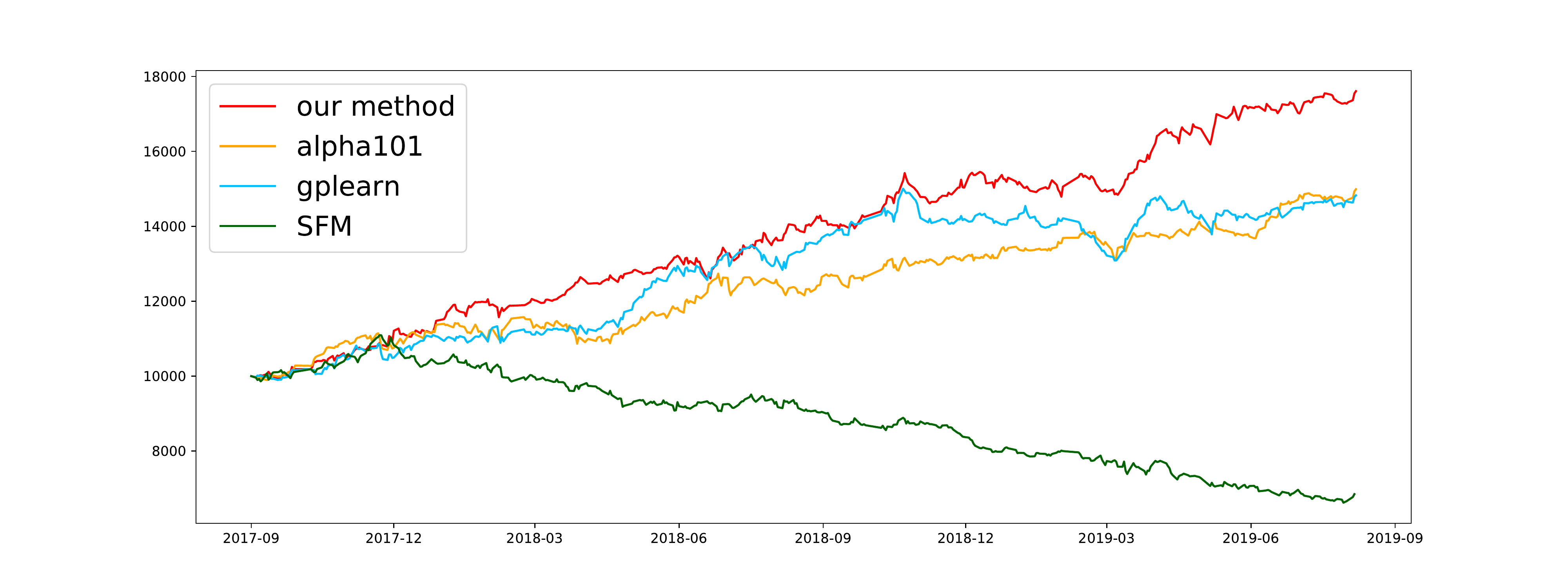} 
  } 
  \caption{The Profit\&Loss graph of backtests.} 
  \label{fig:back-testing} %% label for entire Figure 
  \setlength{\belowcaptionskip}{-3cm}
\end{figure}

\begin{figure}[H] 
  \centering
  \includegraphics[height=3cm, width=8cm]{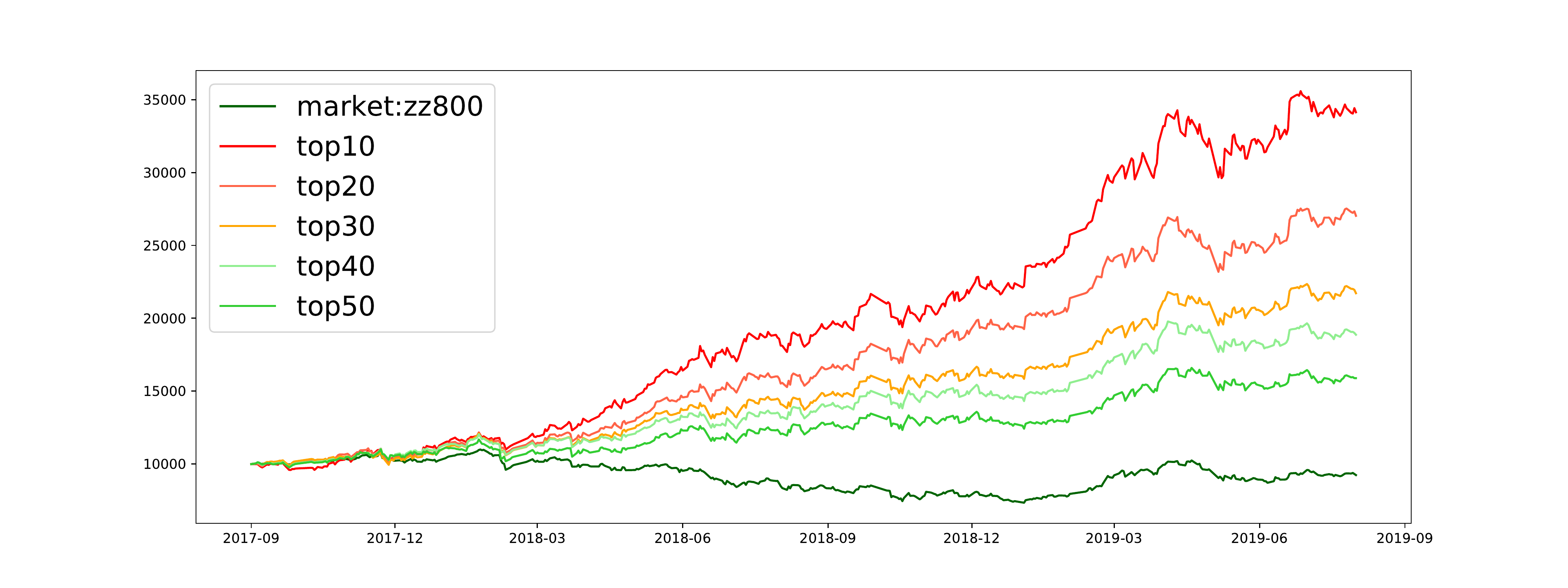} 
  \caption{Comparison on backtesting strategies (longing \textbf{top 10, top 20, ..., top50}). Holding period $h=1$.} 
  \label{fig:back-testing_top} %% label for entire Figure 
  \setlength{\belowcaptionskip}{-3cm}
\end{figure}

For each holding period, we collect the top 150 formulaic alphas generated by {\em AutoAlpha} which have the highest IC in the training period for model training. Then we use the trained model to predict the stock rankings in the testing period. We ensure the overall procedure does not use future information that is not available at the trading time. We perform backtests using historical data for holding periods $h=1$ and $5$. For a given holding period $h$, the investor invests in the top $10$ stocks at the close price according to the predicted rankings in each day. Then the stocks are held for $h$ days and sold at the close price at the end of the holding period. The transaction cost is $0.3\%$ as accustomed. The stock pool for trading simulation is the CSI 800 Index (zz800). The backtesting results are shown in the Figure \ref{fig:back-testing}, \ref{fig:back-testing_top} and Table \ref{tab:back-testing}. 

% \begin{figure}[H] 
%   \centering 

%   \caption{The Profit\&Loss graph relative to the market.} 
%   \label{fig:back-testing_relative} %% label for entire Figure 
%   \setlength{\belowcaptionskip}{-3cm}
% \end{figure}

\section{Related Work}
% 第一部分写历史上的model
% 第二部分写alpha 101，引用它的文章，以及feature extraction
% 第三部分写GA相关

Data mining and machine learning techniques have been used extensively to address a variety of problems in finance. In particular, our work is closely related to feature extraction in machine learning  (see e.g., \cite{guyon2008feature}). Feature extraction has been widely adopted in financial prediction. \cite{zhang2018stockassistant} extracted features from the semantic information in stock comment text for reliability modeling of stock comments. \cite{huang2008integrating} proposed a GA-based model to extract wavelet features for stock index prediction. Our work develops a substantially different GA-based algorithm and extracts general formulaic alpha factors.

How to maintain diversity has always been a critical issue in genetic algorithms \cite{gupta2012overview}. The methods to encourage diversity includes Nitching \cite{sareni1998fitness}, Crowding \cite{mahfoud1992crowding}, Sharing \cite{goldberg1987genetic}, etc. The replacement method we use derives from the Steady State Genetic Algorithms (SSGAs) \cite{engelbrecht2007computational}. A number of replacement methods have been developed for SSGAs, including the parent-offspring competition we use \cite{smith1999replacement}. The Quality Diversity (QD) is designed to illuminate the diverse individuals with high performance. There are many quality diversity algorithms designated for different kinds of problems \cite{pugh2016quality}. Common examples include Novelty Search with Local Competition (NSLC) \cite{lehman2011evolving} and Multi-dimensional Archive of Phenotypic Elites (MAP-Elites) \cite{mouret2015illuminating}.

\section{Conclusions}
In this paper, we propose {\em AutoAlpha},
an efficient algorithm that automatically discovers
effective and diverse formulaic alphas for quantitative investment. We first propose a hierarchical structure to quickly locate the promising space for search. Secondly, we propose a new {\em PCA-QD} search to guide the search away from the explored areas. Thirdly, we utilize the warm start method and the parent-offspring replacement method to prevent the premature convergence problem. The backtests and comparisons with several baselines show the effectiveness of our method. Finally, we remark that {\em AutoAlpha} can be also viewed as an approach for automatic feature extraction. As the market becomes more efficient, discovering alpha factors becomes more difficult and automatically extracting effective features is a promising future direction for quantitative investment.

% \textbf{The second} is to incorporate human knowledge in the search. Right now, after manually screened by financial engineers (based on the stability and economic/financial meaning), several alphas discovered by {\em AutoAlpha} has been deployed in real life trading. In fact, manual selection can further improve the results. It would be very desirable to incorporate human knowledge to generate more meaningful and robust alpha factors.
\newpage
\bibliographystyle{named}
\bibliography{ijcai20}

\end{document}